# An Efficient Method for Uncertainty Propagation in Robust Software Performance Estimation

Aldeida Aleti[1], Catia Trubiani[b], André van Hoorn[c], Pooyan Jamshidi[d]

[a]*Faculty of Information Technology, Monash University, 900 Dandenong Road Caulfield East, VIC, 3145, Australia*
[b]*Gran Sasso Science Institute, Viale Francesco Crispi 7, 67100 LAquila, Italy*
[c]*Institute of Software Technology, Universität Stuttgart, Universitätsstraße 38 D-70569, Stuttgart, Germany*
[d]*School of Computer Science, Carnegie Mellon University, 5000 Forbes Avenue, Pittsburgh, PA 15213, USA*

**Abstract**

Software engineers often have to estimate the performance of a software system before having full knowledge of the system parameters, such as workload and operational profile. These uncertain parameters inevitably affect the accuracy of quality evaluations, and the ability to judge if the system can continue to fulfil performance requirements if parameter results are different from expected. Previous work has addressed this problem by modelling the potential values of uncertain parameters as probability distribution functions, and estimating the robustness of the system using Monte Carlo-based methods. These approaches require a large number of samples, which results in high computational cost and long waiting times.

To address the computational inefficiency of existing approaches, we employ Polynomial Chaos Expansion (PCE) as a rigorous method for uncertainty propagation and further extend its use to robust performance estimation. The aim is to assess if the software system is robust, i.e., it can withstand possible changes in parameter values, and continue to meet performance requirements. PCE is a very efficient technique, and requires significantly less computations to accurately estimate the distribution of performance indices. Through three very different case studies from different phases of software development and heterogeneous application domains, we show that PCE can accurately (>97%) estimate the robustness of various performance indices, and saves up to 225 hours of performance evaluation time when compared to Monte Carlo Simulation.

---

*Corresponding author
URL:* `aldeida.aleti@monash.edu` (Aldeida Aleti)

# 1. Introduction

Performance is a crucial quality attribute of a software system that can be estimated at different stages of its life-cycle, from the architectural design phase to when the system is implemented and running [1]. When the performance is measured early in the development process, the aim is to derive the appropriate architectural decisions that improve the performance of the system. At this stage, performance models (e.g., Petri Nets [2], Queueing Networks [3], Layered Queueing Networks (LQN) [4]) are used to describe how system operations use resources, and how resource contention affects operations. Decisions made at early development phases greatly impact the quality of the final software product, and wrong decisions may imply an expensive rework, possibly involving the overall software system. Discovering performance issues early in the software development process may help avoid expensive fixes after the system has been built. At later stages of the software's lifecycle, performance models rely on actual measurements, and have additional uses, in particular: (i) design of performance tests; (ii) configuration of products for delivery; (iii) evaluation of planned evolutions of the design, while recognising that no system is ever final [5].

Performance estimationat early stages of software development is difficult, as there are many aspects that may be unknown or uncertain, such as the design decisions, code, and execution environment. In many application domains, such as heterogeneous distributed systems, enterprise applications, and cloud computing, performance evaluation is also affected by external factors, such as increasingly fluctuating workloads and changing scenarios [6, 7]. As a result, the software engineer is often forced to make decisions under uncertainty (i.e., related to the design or refactoring of the system), without knowing the exact impact of those decisions on performance.

A possible solution is to explicitly define the uncertainty in input parameters during the performance estimation process [8]. This allows for approximations of performance indices in cases where the system behaviour can not be determined or precisely evaluated in advance. In our previous work [9], we introduced a probabilistic formulation of input parameter uncertainties. The approach samples the distribution of the uncertain input parameters using a Monte Carlo-based approach, and solves a queuing network for each sample to evaluate the respective performance values. The histogram of output values is then used to evaluate how confident one should be that the system performance would not go below or above a certain required value.

While the Monte Carlo-based method provides a useful estimate of the probability distribution for the performance indices when uncertain parameters are present, it requires a large number of samples and very high computational cost, which makes it computationally expensive for design-time analytic solutions or on-line performance evaluations, and inconvenient for runtime applications.

To address this issue, we introduce a new approach for estimating performance robustness, which is based on polynomial chaos expansion (PCE) [10]. PCE is a stochastic response surface model that computes cumulative distribution functions for performance indices affected by uncertain input parameters. Accurate statistics can be obtained with only a few runs of computationally expensive queuing network solutions or measurements from running systems.

The proposed technique is applied to three very different datasets evaluated at various phases of software development and from heterogeneous application domains. These are:

*E-commerce* system, which is modelled at design-time with Layered Queueing Networks (LQN) [11], and two real systems, *JPetStore*, which is a Java enterprise application running in a controlled environment under load generation, and *WordCount*, a Stream Processing System (SPS) benchmark implemented with Apache Storm and running on a cloud cluster with OpenNebula. Experimental results demonstrate the effectiveness, accuracy, and robustness of our approach. The PCE method produces very low errors even with only 100 samples, saving up to 225 hours when compared to the Monte Carlo Simulation approach. The robustness of the technique was tested on noisy data with different levels of signal-to-noise ratios. The PCE method was at least 80% accurate in propagating uncertainty, with accuracy above 97% in the majority of the cases affected by noise.

## 2. Related Work

Issues arising from uncertain parameter inputs that affect the robustness of a software system have been investigated in the literature, with a main focus on quality attributes, such as reliability [12, 13] and performance [9]. Here, we review papers that are most relevant to our approach, by categorising them in terms of methods that use machine learning techniques for performance prediction, approaches for uncertainty analysis, and more in general techniques that can be used for uncertainty quantification and propagation.

*2.1. Performance Estimation with Machine Learning*

As performance evaluations and measurements can be quite costly, there has been a large body of work that has focused on reducing the number of evaluations required to make accurate estimations. To this end, several learning techniques have proven to be quite effective. Minimising the learning data when training these machine learning models from real observations, such as average arrival rates and response times [14, 15] has been a research priority.

The response surface methodology (RSM) explores the relationships between several explanatory variables and one or more response variables, in particular a sequence of designed experiments is used to obtain an optimal response [16, 17]. Such methodology has been applied for reliability-based design optimisation in [18], for multidisciplinary design optimisation [19, 20], and robust optimisation [21, 22]. A survey of sampling-based methods for uncertainty and sensitivity analysis is provided in [23], and a survey of fitness approximation methods applied in evolutionary computation is reported in [24].

When considering the performance indices of software systems, the type of applied machine learning techniques resulted to be quite diverse. For instance, Courtois and Woodside [25] use regression functions to derive performance models from measured data. An iterative process incorporates new measurements until the model's accuracy is considered high enough. In [26], service resource consumption is derived from an optimized set of response time measurements. Instead, Wang et al. [27] use Bayesian inference and Gibbs sampling to model service demand from queue length data. Kattepur and Nambiar [28] employ spline interpolation of service demands to propose a modified version of the multi-tiered web applications algorithm that provides more accurate estimates of maximum throughput and response time.

Machine learning approaches have been also proposed for performance prediction of highly configurable systems [29, 30, 31, 32]. The system is considered a black-box model, which is learned or deduced from performance measurements of sampled configurations. A common strategy is to execute the system in different configurations and use machine learning techniques to generalize a model that characterize system performance [33].

Zheng et al. [34] applied Kalman filters to update parameters of queueing models as the software system evolves. Ghanbari et al. [35] derived performance parameters (such as service demands) by a combination of K-means clustering algorithm with a tracking filter for grouping classes of services. In [36] statistical techniques are applied to tune the application parameter space, and two classes of predictive models, piecewise polynomial regression and artificial neural networks are compared. Sharma et al. [37] apply independent component analysis to categorize workload requests and to identify their resource demands using measurement results such as CPU and network usage. Finally, genetic programming [38] has been used to derive software performance curves, but the evaluation is restricted to a limited set of input parameters that vary in a narrow interval.

Our approach differs from these works because we consider uncertain parameters of different nature (spanning from workload, operational profile, software resource demand, and hardware service time), and the values of such uncertain parameters can be expressed in different forms, such as probability distribution functions, intervals, random seeds, etc. This provides greater flexibility for software engineers in modelling parameters of different characteristics. In literature ML techniques are well-know to require a substantial amount of training data in order to provide accurate performance estimations. PCE, on the other hand, is a cost effective method that requires significantly less training data to estimate the performance of software systems [39], as also demonstrated by our experiments in this paper. Furthermore, the PCE approach has advantages over other ML methods due to the fact that the mean and variance of the output required for the quantifying the robustness of the system are available in closed form [40].

*2.2. Performance Uncertainty Analysis*

There are several works in the literature that optimise QoS properties of software systems affected by variability in system or environment parameters. For instance, Menasce et al. [41] deal with uncertainties by reconfiguring software architectures at runtime. Kowal et al. [42] leverage the commonalities across variants of software systems to analyse uncertainty in terms of both parametric changes (which affect the values of performance annotations) and structural changes (which may affect the topology of performance models). Incerto et al. [43] express uncertainty as symbols in the specification of queuing network performance models and derive actual values of symbols by means of satisfiability modulo theory. Perez-Palacin et al. [44] proposed a methodology to guide software engineers in the process of recognizing and managing the existence of uncertainty. Jamshidi et al. [45, 7, 46] deal with uncertainty of workload in dynamic and uncertain cloud environments [47].

Previous work has tackled the problem of model-based performance and reliability evaluation of software systems in presence of uncertainties [9, 12, 48, 49]. The probabilistic formulation of parameter uncertainties is sampled using a Monte Carlo-based approach to systematically assess the robustness of a software system under uncertainty.

As drawback, Monte Carlo simulation requires the creation of a large number of samples, which can be very computationally expensive [50], as the performance and reliability models have to be re-evaluated for each sample.

Esfahani et al. [6] introduce GuideArch, which employs fuzzy mathematical methods to reason about uncertainty. In GuideArch, the impact of architecture design alternatives on software properties is modelled as a triangular fuzzy value. The method enables the ranking of software architectures, finding the optimal architecture, and identifying critical design decisions.

Another approach that provides a quantitative method for comparing different design alternatives dealing with uncertainty is rArcheopterix (robust ArcheOpterix) [48, 12, 13]. The rArcheopterix framework considers the uncertainty of design-time parameter estimates and optimises embedded system architectures for clearly defined robust quality goals, such as confidence intervals.

Similarly to rArcheOpterix [48], we consider uncertain parameters as probability distributions. However, rArcheOpterix uses Monte Carlo-based method, which is computationally expensive. In this paper, we propose the PCE approach, which significantly reduces the number of samples required to perform uncertainty propagation.

*2.3. Uncertainty Modelling and Propagation*

The related work in uncertainty modelling and propagation aims at measuring the impact of uncertain input parameters on the system output. One way of classifying these methods is based on how uncertainty is described. The main categories are:

- **Probabilistic approaches**, which assume that the probability density functions (PDFs) of the uncertain parameters are known. These methods are known as parametric approaches and include solving techniques such as (i) maximum likelihood [51], (ii) method of moments [52], and (iii) Bayesian estimation [52]. In order to use these approaches, one has to choose the form of the model, e.g., Gaussian, and determine the coefficients of that model.

- **Possibilistic approaches** are methods that use a membership function to model uncertain input parameters. These methods are also known as non-parametric or fuzzy approaches, originally introduced by Zadeh [53]. Uncertain parameters are described using linguistic categories with fuzzy boundaries.

- **Information gap decision theory** [54] estimates the impact of uncertain parameters by the deviation of errors, i.e., the difference between the parameters and their estimation. It is usually used in cases historical data is not available, and uncertain parameters cannot be described with probability distributions functions or membership functions.

- **Interval analysis** was first introduced by Moore [55] and models uncertain parameters as intervals. Differently from probabilistic approaches, interval analysis is similar to a worst-case analysis. It aims to investigate the limits of variation, without assumptions of distribution.

Our approach belongs to the probabilistic kind, and is applied for the first time to model and propagate uncertainty in software systems.

# 3. Polynomial Chaos Expansion for Uncertainty Propagation in Robust Performance Estimation

Any software system, whether it is at the modelling stage or while running, has uncertainties associated with it. For instance, web-based systems would have to deal with an uncertain number of users, who may generate an uncertain number of requests for the various services that the system offers. At the modelling stage, a software architect would not have a full picture of the environment in which the software would be deployed. All these uncertain parameters affect the quality attributes of the system, and most importantly, performance.

It is possible to estimate the range or distribution of uncertain parameters from historical data, or using expert knowledge. It is yet unclear, however, how to estimate the performance (e.g., response time and resource utilization) of a software system when it is subject to uncertain parameter values. In this work, we seek to develop a solution to this problem.

As software systems usually have a large number of uncertain parameters with many possible values, it becomes infeasible to run exhaustive evaluations of the sample space and all possible value combinations. Instead, the method that we propose to use for estimating performance robustness explicitly considers parametric uncertainties and uses polynomial expansions to propagate these uncertainties, which makes it computationally more efficient than the standard Monte Carlo-based schemes that have previously been used in this area (please refer to Section 2).

*3.1. Polynomial Chaos Expansion*

Given an uncertain parameter $x$ (e.g., number of users for a service), and a performance index (e.g., response time of the service) $y$, polynomial chaos expansion estimates how the output varies as the input varies, thus propagating the uncertainty in the input into the output.

As the $x$ parameter is uncertain, it represents a random variable $X$, and it is specified via a specific random variable $\xi$ and its PDF $\rho(\xi)$, as follows:

$$X = f(\xi) \qquad (1)$$

The random variable $\xi$ can be uniform, exponential, etc., depending on the modelling choice. We seek for an appropriate function $f$ such that given $\rho(\xi)$, we can achieve the required distribution of $X$. PCE takes a representation of an uncertain input (eq. 1), and expands the function $f$ in terms of a set of orthogonal polynomials.

The polynomial basis comprises polynomials $\psi_0 = 1, \psi_1, \psi_2, ...$, where $\psi_i$ is a polynomial of order $i$, and they satisfy the orthogonality condition:

$$\langle \psi_i, \psi_j \rangle = \int \psi_i(\xi)\psi_j(\xi)\rho(\xi)d\xi = 0, \quad \forall i \neq j \qquad (2)$$

which means that the covariance between any two different $\psi_i(\xi)$ is zero, that is they are uncorrelated.

Using the orthogonal basis polynomials, we can write:

$$X = f(\xi) = \sum_{i=0}^{\infty} a_i \psi_i(\xi), \qquad (3)$$

where $a_i$ are the expansion coefficients (the mode strength). The combination of $a_i$ and $\psi_i$ is called the $i$-th mode, for which a unique expansion exists where the mode strengths are given by

$$a = \langle f, \psi_i \rangle / \langle \psi_i, \psi_i \rangle. \tag{4}$$

The denominator $\langle \psi_i, \psi_i \rangle$ is computed from the orthogonal polynomials, whereas the enumerator $\langle f, \psi_i \rangle$ is an integral of the form given by eq. 2, which is evaluated numerically for complex forms of $f$.

Any expansion of $\xi$ in the form of eq. 3 is called PCE. There are many possible functions $f$ for a given $X$ and $\xi$ distribution, hence there can be many possible PCEs of a given $X$ using a given random variable $\xi$.

For practical reasons, PCEs are usually truncated to a finite number of terms $d$ as follows:

$$X = f(\xi) \approx \sum_{i=0}^{d} a_i \psi_i(\xi), \tag{5}$$

The term $d$ should be large enough for the approximation to be accurate. In practice, a good representation $f$ is one for which the truncated representation $f_d$ with small $d$ will be accurate. Since polynomials are orthogonal, they can be written uniquely as an expansion in the orthogonal family, which means that coefficients $a_i$ are well defined. The polynomials are chosen based on the distribution of the chosen random variable $\xi$. The polynomial types used for standard forms of continuous probability distributions are listed in Table 1.

Table 1: Askey scheme of continuous hyper-geometric polynomials used for standard forms of continuous probability distributions.

| Distribution | Density function | Polynomial | Weight function | Range |
|---|---|---|---|---|
| Normal | $\frac{1}{\sqrt{2\pi}} e^{\frac{-\xi^2}{2}}$ | Hermite | $e^{\frac{-\xi^2}{2}}$ | $[-\infty, \infty]$ |
| Uniform | $\frac{1}{2}$ | Legendre | 1 | $[-1, 1]$ |
| Beta | $\frac{(1-\xi)^\alpha (1+\xi)^\beta}{2^{\alpha+\beta+1} B(\alpha+1, \beta+1)}$ | Jacobi | $(1-\xi)^\alpha (1+\xi)^\beta$ | $[-1, 1]$ |
| Exponential | $e^{-\xi}$ | Laguerre | $e^{-\xi}$ | $[0, \infty]$ |
| Gamma | $\frac{\xi^\alpha e^{-\xi}}{\Gamma(\alpha+1)}$ | Generalised Laguerre | $x^\alpha e^{-\xi}$ | $[0, \infty]$ |

The PDF of $\xi$ can take any form, e.g., discrete, continuous, discretised continuous, specified analytically, using histograms, datasets, or moments. The key point in PC theory is that the polynomial basis is linked with the distribution of $\xi$, which dictates the polynomial basis functions. The weighting function of the orthogonal polynomials in Table 1 is identical to the probability density function of the corresponding input distribution, which dictates the choice of the appropriate polynomial.

Usually, real-world systems have more than one uncertain input parameter. In this case, $X$, $\xi$, and the coefficients (mode strengths) $a_i$ are vectors, hence we write them as $\mathbf{X}$, $\Xi$, and $\mathbf{a}_i$. It follows that the polynomial $\psi_i(\Xi)$ is multivariate, hence we can write it as $\Psi_i$.

If the random variables $\Xi$ are independent, $\Psi_i$ is a tensor product of the polynomial bases for each $\Xi_i$. More specifically, if $\Xi$ comprises $m$ id random variables, and given index $\mathbf{j}=(j_1, j_2, ..., j_m)$, we can write:

$$\Psi_j(\Xi) = \prod_{i=1}^{m} \psi_{j_i}(\Xi_i) \qquad (6)$$

In case the input variables are not independent, they can be transformed using Independent Component Analysis (ICA) to remove any correlations [56]. ICA is a source separation method which aims at estimating independent unobservable (latent) variables which are mixed into observed variables. The main idea of the method is in its search for non-Gaussian components, which rely on the covariance matrix of random variables, and combines a static linear mixture model with higher order statistics in order to identify unobservable variables. An example is the variation of service demands due to changes in caching effectiveness. This occurs when processor workload increases, or there exists competing workload demands, affecting the number of user requests. As a result, these parameters will be correlated.

*3.2. Uncertainty Propagation*

The primary objective of this work is to create a PCE expansion for performance indices given uncertain input parameters. Assume the output of a software system (performance index) is $y$ when given an input $x$, i.e., $y = \eta(x)$, where $\eta$ represents the performance of the software system. As in the previous section, we assume this input is uncertain, hence it is represented as a random variable $X$. As a result, the output is a random variable $Y = \eta(X)$. Both $X$ and $Y$ can be vectors, if there are more than one input and one output.

Assume that we have a PCE for $X$ as described in eq. 5, and we seek to represent the output with another PCE as follows:

$$Y = g(\Xi) = \sum_{i=0}^{\infty} b_i \psi_i(\Xi), \qquad (7)$$

and in the truncated form as

$$Y = g(\Xi) \approx \sum_{i=0}^{d} b_i \psi_i(\Xi). \qquad (8)$$

The random variable $\Xi$, the mode functions $\psi_i$, and the truncation level $d$ for the PCE of $Y$ are the same as the ones used for the PCE of $X$.

The PC analysis is completed by finding the mode strengths $b_i$, using intrusive or non-intrusive methods. Intrusive methods, such as the stochastic Galerkin technique [40], manipulate the governing equations, and provide semi-analytical solutions for uncertainty analysis. These methods are complex and computationally expensive, since they require symbolic manipulations, hence we use a non-intrusive and more efficient method, i.e., sparse quadrature [57].

Non-intrusive methods treat the system as a black box, and solve the following equation:

$$\sum_{i=0}^{d} b_i \psi_i(\xi) = \eta(f_d(\xi)), \tag{9}$$

where $f_d(\xi) = \sum_{i=0}^{d} a_i \psi_i(\xi)$, by evaluation the system at different values of the random variable $\xi$.

Once the mode strengths have been identified, and the PCE for the output $y = \sum_{i=0}^{d} b_i \psi_i(\xi)$ has been established, the distribution $Y$ as induced by distribution $X$ is estimated.

Next, characteristic statistical properties of $Y$ useful for uncertainty quantification, such as mean and variance can be exactly calculated. The first two moments of $Y$ are obtained using properties of the orthogonal polynomials:

$$E(Y) = b_0, \quad E(Y^2) = \sum_{i=0}^{d} b_i^2 \langle \psi_i, \psi_i \rangle. \tag{10}$$

from which the variance of $Y$ can easily be calculated. Since for orthogonal polynomials $\langle \psi_i, psi_i \rangle = 1$, the variance is equal to

$$E(Y^2) = \sum_{i=0}^{d} b_i^2. \tag{11}$$

*3.3. Estimating Software Performance Robustness with PCE*

The PCE method is applied to estimate the performance of software systems under uncertain input parameters. The main steps of the method are depicted in Algorithm 1 using the parameters from the running example of this section.

---
**Algorithm 1** PCE for Robust Performance Estimation.
---
1: **procedure** PCE
2:     $u \approx \sum_{i=0}^{d} a_i \psi_i(\xi)$        ▷ *Number of users is represented using random variable $\xi$.*
3:     $rt \approx \sum_{i=0}^{d} b_i \psi_i(\xi)$        ▷ *Response time is represented in terms of $\xi$.*
4:     $a = \langle \sum_{i=0}^{d} a_i \psi_i(\xi), \psi_i \rangle / \langle \psi_i, \psi_i \rangle$        ▷ *Calculate the mode strengths a.*
5:     $b = \langle \sum_{i=0}^{d} b_i \psi_i(\xi), \psi_i \rangle / \langle \psi_i, \psi_i \rangle$        ▷ *Calculate the mode strengths b.*
6:     $E(rt) = b_0$        ▷ *Calculate the mean of response time.*
7:     $E(rt^2) = \sum_{i=0}^{d} b_i^2$        ▷ *Calculate the variance of response time.*
8:     CoV$= \frac{\sqrt{E(rt^2)}}{E(rt)}$        ▷ *Calculate the coefficient of variation of response time.*
9: **end procedure**
---

The first step of this process (line 2 of Algorithm 1) is concerned with representing uncertain input parameters in terms of random variable defined by their probability distributions. Depending on the type of the uncertain input parameter, different probability density functions may be appropriate.

For instance, consider a web-based system that provides certain services, and we are interested in estimating whether the system is robust in terms of response time, i.e., we would like to make sure that our system can meet response time requirements of our users despite the uncertain parameters that may affect it. One potential uncertain parameter is the number of users. Through historical data (e.g., considering how the system has been used in the past), assume we observe that this parameter follows a normal distribution with mean $\mu = 100$ and standard deviation $\sigma = 20$.

As described in line 2 of Algorithm 1, the number of users $u$ is expressed in terms of a standard normal random variable $\xi$ in $N(0, 1)$. Given $X$, the aim is to estimate how this uncertainty manifests itself in the performance indices of the system, such as the response time. To this end, the response time $trt$ is expressed in terms of the same random variable $\xi$ as follows:

$$rt \approx \sum_{i=0}^{d} b_i \psi_i(\xi) \qquad (12)$$

The coefficients $b_i$ are calculated using the sparse quadrature method [57], as illustrated in line 4 and 5 of Algorithm 1. In our example, this represents the response surface of response time as calculated by PCE, which can be used for performance robustness quantification.

Performance robustness is defined as either the lack or low level of performance variation in response to a perturbation. To illustrate this concept, the output of the PCE model is presented in Figure 1 for two different software systems that are affected by the same uncertain parameter (number of users) with mean 100 and standard deviation 20.

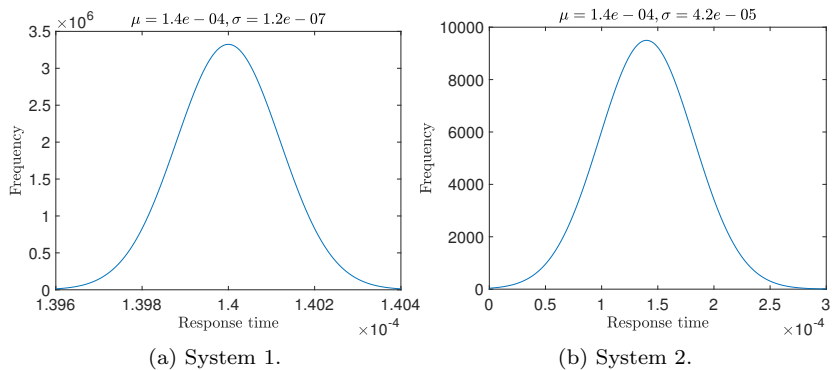

Figure 1: Distribution of potential number of users estimated using 1,000 samples.

The response time of both systems has the same mean, but differs in standard deviation: the response time of System 1 shown in Figure 1a has a lower standard deviation ($\sigma = 1.2e - 07$) than the response time of System 2 shown in Figure 1b ($\sigma = 4.2e - 05$). Standard deviation is a commonly used measure of variation and robustness. This metric can be estimated using Equation 10. However, standard deviation, while it measures dispersion, does not allow for comparing how robust a system is for different performance indices coming from different distribution, as they may vary greatly in the means about which they occur.

In this paper, we propose to use the coefficient of variation (CoV), which can be calculated by taking the ratio of mean and standard deviation, and as a result cancelling the units.

$$\text{CoV} = \frac{\sqrt{E(rt^2)}}{E(rt)} \tag{13}$$

Both the mean (first moment) $E(rt)$ and the variance (second moment) $E(rt^2)$ are calculated using Equation 10. CoV is expressed as a percentage, and estimates the variability of the performance indices with respect to the mean, as reported in Equation 13 and line 4 of Algorithm 1. The means tell us whether the systems meet performance requirements, whereas the CoVs indicate whether these means are robust. A system with low CoV is less dispersed, hence more robust towards the variation of the input parameters. In the example presented in Figure 1, CoV of System 1 is 0.08%, whereas the CoV of System 2 is 30%, hence it can be concluded that the response time of System 1 is more robust than the response time of System 2, and would be less influenced by variations in the number of users.

## 4. Experimental Evaluation

To evaluate the PCE technique for robust software performance estimation, we performed an experimental study on three different software systems. The case studies are chosen such that they represent different phases of software development, and have different numbers of uncertain input parameters with various distributions. The nature of the systems is also different with uncertainty levels of different scales. Each run has a fixed set of parameters that does not vary during the experiment.

We empirically estimate the ranges and distribution of uncertain parameters using experimental measurements of the system. The probability density function is estimated from the data. While there are different ways of estimating the probability density function, a typical approach is using kernel density [58]. A kernel distribution produces a nonparametric probability density estimate that fits to the data. The resulting distribution is defined by a kernel density estimator, a smoothing function that determines the shape of the curve used to generate the probability density function, and a bandwidth value that determines the range.

We use cross-validation for model selection, while the accuracy of the selected degree for the PCE is measured by the leave-one-out error [59]. The method consists of running the PCE analysis on $N$ sets of a reduced experimental design selected from the available data, and comparing its prediction on the excluded points with the real output values. The generalisation error is estimated for each set of reduced experimental design, and averaged over $N$ sets. This method has been shown to be more reliable and to overcome any issue with over-fitting present in other approaches [60].

We also investigate the efficiency of the method by using different numbers of samples in PCE model construction, and robustness, which is determined by introducing different levels of noise in measurements. Finally, we evaluate how PCE performs when compared to the commonly used Monte Carlo approach. Different samples were used for PCE and the Monte Carlo Method. Results are shown for each case study individually, and summarised at the end.

*4.1. Benchmark Method – Monte Carlo Simulation*

We compare the efficiency of the PCE method against a commonly used approach: Monte Carlo Simulation [13]. Monte Carlo-based methods usually evaluate a large number of samples that consider various combinations of uncertain parameters. As in the PCE approach, uncertain parameters are sampled from their respective probability distribution, and performance indices are measured as described in Algorithm 2. The MC method is terminated when the desired statistical significance is achieved, measured by the relative error $e$ in line 7 of Algorithm 2.

---
**Algorithm 2** Monte-Carlo method with dynamic stopping criterion.
---
1: **procedure** MONTECARLO( )
2: $\quad i = 1$
3: $\quad$ **while** $e > 0.05$ **do** $\quad\triangleright$ Tolerance level is set at 0.05 for 95% confidence.
4: $\quad\quad r_i =$ MONTECARLOSIMULATION(samples) $\triangleright$ Perform Monte Carlo Simulation.
5: $\quad\quad \overline{r} = \frac{\sum_{j=0}^{i} r_j}{i}$ $\quad\triangleright$ Calculate the mean over all samples.
6: $\quad\quad \overline{r^2} = \frac{\sum_{j=0}^{i} r_j^2}{i}$ $\quad\triangleright$ Calculate the mean-square over all samples.
7: $\quad\quad e = \frac{2z_{(1-\alpha/2)}}{\sqrt{i}} \frac{\sqrt{\overline{r^2} - \overline{r}^2}}{\overline{r}}$ $\quad\triangleright$ $\alpha$ is the desired significance of the test.
8: $\quad\quad i = i + 1$
9: $\quad$ **end while**
10: **end procedure**
---

The parameter $z$ in line 7 refers to the inverse cumulative density value of the standard normal distribution (for less than 30 samples the t-distribution is used), whereas $\alpha$ is the desired significance of the test. For further details and illustration of the dynamic stopping criterion we refer the interested reader to [13].

## 5. Case Study 1 – E-commerce

The E-commerce system is a web-based application that manages business data related to books and movies: *guest* users can browse catalogues, while *customer* users can make selections of items to be purchased. A Layered Queueing Network (LQN) model for this case study has been used for running model-based performance analysis [61]. Here we introduce uncertainties in the specification of input parameters as described in the following section.

*5.1. Experimental Design*

Table 2 reports the uncertain parameters defined for this case study. Depending on the source of uncertainty, parameters may result in different distributions and range of possible values. Software designers had a colloquium with the system owner, and they agreed to specify the following range of values: (i) workload, which is labelled as *users*, and is uniformly distributed between the range $(49, 98)$; (ii) service demands, denoted as $PAvail$ and $PQual$, which indicate the demand incurred by the services intended to determine the availability and quality of products, respectively; (iii) hardware service times, that are: (1) *DBL*, i.e., light requests to database 25% of time solved in 0.03 UT (unit of time), 50% with 0.015 UT, and 25% with 0.0075 UT, and (2) *DBH* refers to heavy requests to database which are in 50% of the time solved in 0.06 UT (units of time), 25% with 0.009 UT, and 25% with 0.12 UT; (iv) operational profile *DBA* denotes the access to promotions by users, and is triangularly distributed between 0.2 and 0.8, plus 0.5 as mode. The parameters show no correlation between each other, hence are considered independent and no tranformation is required for the PCE. The LQN Solver tool [62] is used to analyze the runtime behaviour of the software system and get measurements for different performance indices representing response time, throughput, and utilization. The evaluation of each sample takes approximately 3 minutes.

Table 2: E-commerce – uncertainty specification.

| Parameter | Distribution | Range of values |
|---|---|---|
| users | uniform | (49, 98) |
| PAvail | normal | (2, 8) |
| PQual | normal | (4, 16) |
| DBL | discrete | (0.03, 0.25, 0.015, 0.5, 0.0075, 0.25) |
| DBH | discrete | (0.06, 0.5, 0.009, 0.25, 0.12, 0.25) |
| DBA | triangular | (0.2, 0.8, 0.5) |

*5.2. Results*

The PCE technique is applied to the E-commerce case study. The probability distribution plots in Figure 2 show the results for all performance indices, both predicted ($Y_{PC}$) and measured ($Y$). For the purpose of this experiment, we considered the following performance indices: mean response time and throughput for the *browseCatalog* and *makePurchase* services, and the utilization of database and library components.

The number of samples used in this case is 1,000. The plots indicate that in all scenarios, the probability distributions of the real and predicted values by PCE are almost identical, indicating that the polynomial chaos expansion technique is very accurate.

The detailed results of the polynomial chaos expansion are shown in Table 3. The second column lists the polynomial degree, which shows the complexity of the model. This is derived through an automated exploration of different degrees up to d=30, which gives the best fit. The mean, standard deviation (SD), and coefficient of variance (CoV) are indicators of solution quality. These are estimated through Equation 10 for each performance index reported in Figure. 2.

The CoVs shown in Table 3 indicate that the solution is relatively robust in terms of response time for *makePurchase* service (around 54%), but has very low robustness with respect to the other performance indices, especially for the throughput of the *browseCatalog* service (159.77%) and the utilization of the *library* component (159.76%).

As this case study shows, while a software system may be robust in terms of some performance index, it may suffer in the robustness of other indices. Hence, these aspects should be considered simultaneously when designing new software systems, or while taking refactoring actions.

The degree of the most accurate expansion depends on the number of samples used to construct the PCE. The errors in Table 4 are an indication of the accuracy of the uncertainty propagation and robustness estimation. The PCE approach was very efficient and produced very low errors even with 100 samples. In the majority of the cases, the accuracy of PCE was above 99%, apart from the thoughput for the *makePurchase* service, where the PCE method had an error of 14% for 100 samples.

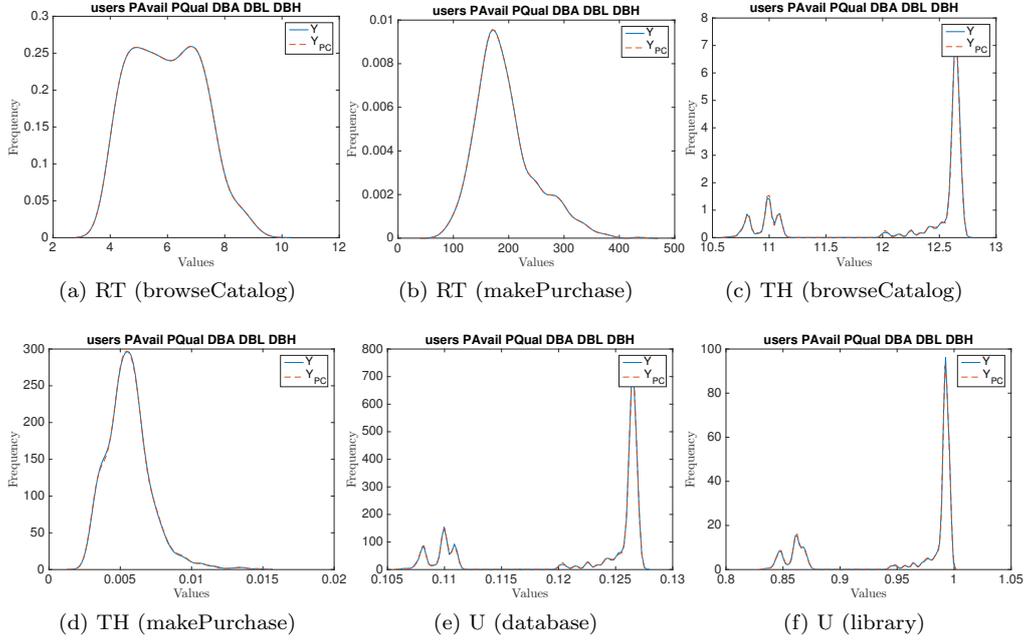

Figure 2: Distribution of the different performance indices for the E-commerce System, as obtained from direct sampling of 1,000 points, as well as from the PCE estimate.

Table 3: PCE results for the E-commerce System.

| Performance | PCE | | | |
| Indices | Degree | Mean | SD | CoV |
|---|---|---|---|---|
| RT (browseCatalog) | 6 | 1.84 | 2.94 | 159% |
| RT (makePurchase) | 8 | 182 | 43.21 | 54% |
| TH (browseCatalog) | 7 | 3.47 | 5.55 | 159.77% |
| RT (makePurchase) | 16 | 0.0062 | 0.0022 | 35.2% |
| U (database) | 6 | 0.038 | 0.057 | 149.55% |
| U (library) | 7 | 0.27 | 0.43 | 159.76% |

For comparison, the last column of Table 4 shows the number of samples required by the Monte Carlo method to accurately estimate the robustness of the system. The confidence value was set to 95%, which means that errors should be less than 5%. As each sample takes approximately 3 minutes to be evaluated, we set the stopping criterion to 1,000 or until the desired level of accuracy has been achieved. In 3 out of 6 scenarios the MC method did not achieve the desired level of accuracy with the allowed 1,000 samples:

Table 4: Accuracy of PCE estimates with different number of samples for the E-commerce System.

| Performance | PCE Samples | | | | MC Samples |
|---|---|---|---|---|---|
| Indices | 1000 | 700 | 300 | 100 | Required |
| RT (browseCatalog) | 5.2 E-05 | 7.2 E-05 | 1.2 E-04 | 2.9 E-04 | >1,000 |
| RT (makePurchase) | 5.1 E-05 | 7.1 E-05 | 1.2 E-04 | 2.9 E-04 | >1,000 |
| TH (browseCatalog) | 2.2 E-04 | 5.72 E-04 | 1.19 E-03 | 3.25 E-03 | 549 |
| TH (makePurchase) | 1.9 E-02 | 4.9 E-02 | 1.1 E-01 | 1.4 E-01 | >1,000 |
| U (database) | 4.5 E-04 | 6.6 E-04 | 4.1 E-04 | 6.2 E-03 | 549 |
| U (library) | 3.1 E-04 | 4.83 E-04 | 3.05 E-04 | 2.7 E-03 | 557 |

RT (browseCatalog), RT (makePurchase), and TH (makePurchase). PCE, on the other hand, produced models for these cases that are at least 98% accurate. in the other three cases, while the MC method converged at around 550 samples, PCE required only 100 samples with an error < 1%. This translates to 450 samples times 3 minutes/sample = 22.5 hours in time savings.

Table 5: Leave-one-out cross-validation errors of PCE with different levels of relative added noise (RAN) for the E-commerce System.

| Performance | RAN level | | | |
|---|---|---|---|---|
| Indices | 20 | 10 | 5 | 1 |
| RT (browseCatalog) | 6.94E-03 | 6.29E-02 | 1.71E-01 | 3.43E-01 |
| RT (makePurchase) | 1.40E-04 | 1.70E-04 | 3.50E-04 | 5.70E-04 |
| TH (browseCatalog) | 2.02E-02 | 1.65E-01 | 3.94E-01 | 6.04E-01 |
| TH (makePurchase) | 2.83E-02 | 4.25E-02 | 8.40E-02 | 1.64E-01 |
| U (database) | 2.07E-02 | 6.24E-02 | 1.80E-01 | 3.81E-01 |
| U (library) | 1.11E-02 | 2.87E-02 | 9.29E-02 | 2.48E-01 |

While the PCE achieved high accuracy for most purposes (more than 99%), a more accurate estimate of the tails of the distribution can have significant implications, especially in the case of robust performance estimation. Moreover, the PCE approach estimates the output distributions at a very low computational cost, giving high accuracy even with a low number of samples. The PCE achieved more than 99% accuracy even with 100 samples in the majority of the performance indices. The numerical experiments confirm that the new methodology is competitive in a wide range of parameters, especially where high accuracy is required.

Finally, we investigate the robustness of the PCE technique by adding independent normally distributed random values [63] to the LQN model-based results. To control the level of the noise that we add to the response value, we use the relative added noise (RAN), which is defined as the ratio of the power of response value and the power of added noise:

$$RAN = \frac{\mu}{\sigma} \qquad (14)$$

where $\mu$ is the mean of the response and $\sigma$ is the standard deviation of the noise. We set different levels of RAN, with higher RAN level indicating less discrepancy and noise. The leave-one-out errors shown in Table 5 indicate that the PCE method is quite robust towards introduced noise. In the majority of the cases, the errors are below 20%, even with RAN level of 5, which is the highest noise level.

## 6. Case Study 2 – JPetStore

JPetStore[1] is a distributed Java EE demo application representing a typical web-based online shopping system for selling pets, where an HTML-based web interface enables to perform typical use cases, such as signing in and off, browsing through the product catalog with categories and products, maintaining a virtual shopping cart, and purchasing an order. A more detailed description of a previous version of the application w.r.t. its use cases can be found in our previous work [64].

### 6.1. Experimental Design

The JPetStore system is exposed to synthetic session-based workload executed by a workload generator, according to the methodology described in our previous work [64]. A constant number of concurrent users (defining the level of *workload intensity*) is performing a sequence of inter-related requests (defining the *navigational profile*) to the system. The workload is closed as a new user session is started only after a previous session has terminated.

Table 6: JPetStore – uncertainty specification.

| Parameter  | Distribution | Range of values |
|------------|--------------|-----------------|
| numUsers   | uniform      | (10, 100)       |
| cpuCatalog | uniform      | (0, 700)        |

Table 6 shows that we have two uncertain parameters: (i) the number of users (numUsers) uniformly varies from 10 to 100; (ii) the CPU demand (cpuCatalog) of the *Show Catalog* service that uniformly varies from 0 to 700. These two parameters are not correlated, hence no transformation is required for the PCE. The range of values has been selected after having executed a series of preparation experiments with the goal to explore the capacity range of the application under analysis.

The remaining parameters have a constant value. The experiments are executed in the Emulab[2] testbed which provides a controllable, predictable, and repeatable environment for performance experiments [65]. Two server nodes are used: one for running the load driver and the other running the JPetStore system. The Emulab node type *pc3000*, which are Dell PowerEdge 2850s machines with a dual-core 64-bit Intel Xeon 3 GHz processor and 2 GB RAM, is used for both nodes with a Ubuntu 12.04 LTS operating system (GNU/Linux kernel 3.2.0-56, 64 bit).

The JPetStore system is deployed to a Jetty Servlet container (version 7.6.10). Apache JMeter is of version 2.9. The initial and maximum heap size for both JMeter and Jetty are set to 1.5 GB. On both machines, Oracle Java version 1.6.0_45 is used. Both nodes are connected via 100 Mbit/s LAN links.

---

[1]http://github.com/mybatis/jpetstore-6
[2]http://emulab.net

Each experiment runs for 15 minutes, including a ramp-up time of 30 seconds at the beginning of the experiment. During ramp-up time, the number of active users is increased linearly until the desired number of users is reached. The performance measurements of the first 230 seconds and last 30 seconds are removed to consider only the steady state of the experiment.

The end-to-end response times for each invocation of user actions are recorded by the workload generator. After each experiment run, aggregate statistics for the response time are computed per action. The aggregate statistics include mean (with 95% confidence interval, CI) and median values. We selected the mean response time and throughput of the *View Product* service (also contained in the user sessions), and the utilization of the server node running the JPetStore system as the performance measures of interest.

### 6.2. Results

The results of the JPetStore case study are shown as probability distribution plots for each performance indicator in Figure 3. It can be observed that the PCE method estimates the distribution of the performance indicators subject to the uncertain input parameters with high accuracy. Similar to the results in the first case study, the real and estimated distributions are almost identical in all cases, which confirms the efficiency of PCE in estimating the performance robustness, and it is an indicator of the internal validity of the method.

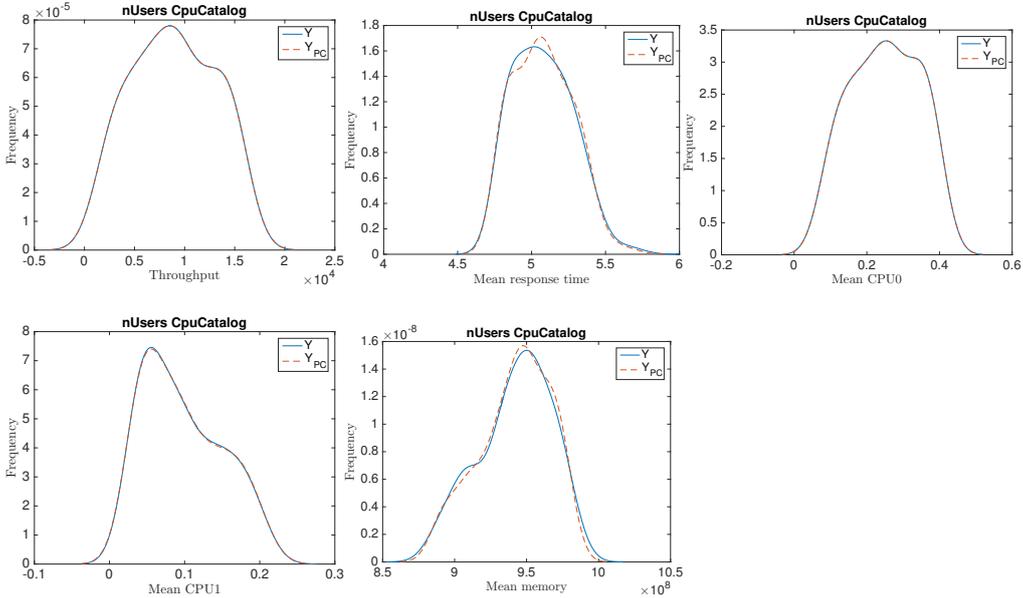

Figure 3: Distribution of the different performance indices for the JPetStore, as obtained from direct sampling of 1,000 points, as well as from the PCE estimate.

The parameters of the constructed PCE model are shown in Table. 7. CoV is the measure that we use to indicate how robust the system is when it is subjected to uncertain input parameters. While the system is quite robust with respect to memory (CoV=3.89%), its robustness is not as high when considering the other performance indicators. One would need to perform refactoring actions to improve the robustness of the system in all performance indicators. This is an optimisation problem, which we intend to investigate in the future.

Table 7: PCE results for JPetStore.

| Performance Indices | PCE | | | |
|---|---|---|---|---|
| | Degree | Mean | SD | CoV |
| Mean throughput | 2 | 8778.16 | 4145.09 | 47.22% |
| Mean response time | 5 | 4.98 | 0.84 | 16.86 % |
| Mean CPU 0 | 14 | 0.25 | 0.09 | 37.97% |
| Mean CPU 1 | 6 | 0.09 | 0.05 | 52.97% |
| Mean memory | 3 | 9.54 E+09 | 3.72 E+08 | 3.89% |

Accuracy of PCE method for JPetStore is investigated for different number of samples and results are shown in Table 8. In all cases, the method performs with high accuracy, with errors lower than 1%, with the highest level of accuracy achieved for throughput. The results are very promising, especially when considering the fact that even with 100 samples, it can produce more than 99% accuracy, which saves in computational cost.

Table 8: Accuracy of the PCE method for JPetStore with different sample sizes.

| Performance Indices | PCE Samples | | | | MC Samples Required |
|---|---|---|---|---|---|
| | 1000 | 700 | 300 | 100 | |
| Throughput | 2.9 E-06 | 2.9 E-06 | 2.9 E-06 | 2.3 E-06 | > 1,000 |
| Response time | 1.7 E-02 | 2.1 E-02 | 4.38 E-02 | 6.1 E-02 | 244 |
| CPU 0 | 8.6 E-04 | 8.8 E-04 | 8.77 E-04 | 1.3 E-03 | > 1,000 |
| CPU 1 | 3.6 E-03 | 3.6 E-03 | 4.29 E-03 | 6.9 E-03 | > 1,000 |
| Memory | 3.9 E-02 | 6.9 E-02 | 7.22 E-02 | 7.1 E-02 | 121 |

The last column of Table 8 presents the number of samples required by the MC method to converge at 95% accuracy. In 3 out of 5 performance indices, the MC method was not able to converge with up to 1,000 samples. Instead, the PCE achieved more than 99% accuracy even with 100 samples. In this case study, each sample takes 15 minutes to simulate, hence the time saved in using 100 samples instead of 1,000 is 225 hours. The MC method's performance is better when modelling the robustness of response time and memory, where only 244 and 121 samples are required. Even in these scenarios, however, the PCE approach requires only 100 samples to achieve high accuracy.

The errors from the PCE modelling were quite low even when noise was introduced, as shown in Table 9, indicating that the method is reliable even with low RAN levels. As expected, the highest errors were observed for RAN level of 1, which ranged from 20%–99%. The highest accuracy was achieved for throughput.

Table 9: Leave-one-out cross-validation errors of PCE with different levels of relative added noise (RAN) for JPetStore.

| Performance Indices | RAN level | | | |
|---|---|---|---|---|
| | **20** | **10** | **5** | **1** |
| Throughput | 2.97 E-06 | 3.01 E-06 | 3.01 E-06 | 3.03 E-06 |
| Response time | 4.46 E-02 | 4.90 E-02 | 6.45 E-02 | 1.13 E-01 |
| CPU 0 | 9.25 E-03 | 2.64 E-02 | 7.61 E-02 | 2.03 E-01 |
| CPU 1 | 9.27 E-03 | 1.94 E-02 | 4.91 E-02 | 1.31 E-01 |
| Memory | 7.20 E-02 | 7.20 E-02 | 7.20 E-02 | 7.20 E-02 |

## 7. Case Study 3 – WordCount

WordCount (cf. Figure 4) is a popular benchmark application for Apache Storm. In WordCount a text file is fed to the system and it counts the number of occurrences of the words in the text file. In Storm, this corresponds to the following operations. A Processing Element (PE) called Spout is responsible to read the input messages (tuples) from a data source (e.g., a Kafka topic) and stream the messages (i.e., sentences) to the topology. Another PE of type Bolt called Splitter is responsible for splitting sentences into words, which are then counted by another PE called Counter. As depicted in Figure 4, WordCount features a three-layer architecture that involves interactions with technologies such as Apache Kafka as data source and a storage to persist the (*word, count*) pairs. This case study is representative of highly configurable systems [33, 30, 66] where the performance is studied in terms of changes in the configuration options while the environmental parameters are kept constant.

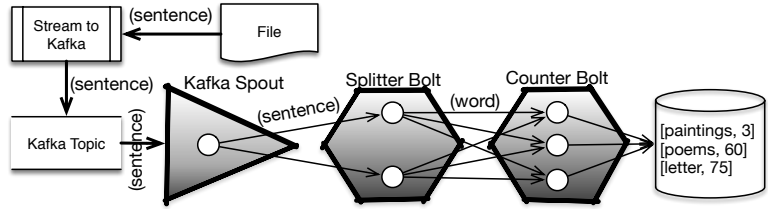

Figure 4: WordCount topology architecture.

### 7.1. Experimental Design

The performance statistics regarding each specific configuration has been collected over a window of 5 minutes (excluding the first two minutes of warm up and the last minute of cluster cleaning between two consequential experiments). The first two minutes are excluded because the monitoring data are not stationary, while the last minute is the time given to the topology to fully process all messages. We then shut down the topology, clean the cluster, and move on to the next experiment.

We replicated each trial for 30 times in order to report the comparison results, and conducted each individual test on a clean cluster, meaning that we killed all processes from a previous run on all cluster nodes before starting the new ones. In addition, we deleted all data generated by the previous run from the storage share between the worker nodes. We ran the experiment on a cluster with 5 nodes (1 Nimbus, 1 ZooKeeper, 3 Supervisor, each 1 CPU, 4GB RAM) on OpenNebula. More details regarding the experiment and raw data are available in [45]. The configuration parameters with their associated values are listed in Table 10.

The range for the parameter values was chosen based on preliminary trials. This setting resulted in a dataset of size 1,404. *spout_wait* is a configuration option that set the time in ms for the *SleepEmptyEmitStrategy* that aims to make the spout in sleeping mode. In other words, this setting will make the spout component to sleep after pushing data, so topologies' throughput will likely decrease as the value of this configuration increases, and this is typically used for internal queuing. Parameters *splitters* and *counters* configuration options determine the level of parallelism of their associated components. The evaluation of each sample takes approximately 8 minutes. As this is a factorial design, we assume no correlation between the uncertain parameters.

### 7.2. Results

Figure 5 presents the entire PDF plots of the throughput and latency computed based on the frequency histograms of the real values $Y$ and the values produced by the PC model $Y_{PC}$. Results are presented for each performance index to test whether the throughput and latency models built using polynomial chaos expansion represent the actual behaviour of the corresponding performance index.

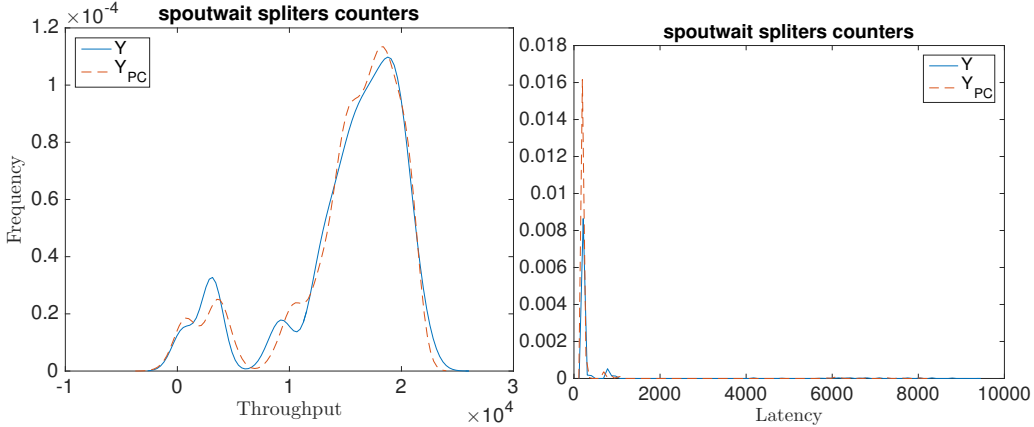

Figure 5: Distribution of the different performance indices for the WordCount system, as obtained from direct sampling of 1,000 points, as well as from the PCE estimate.

Table 10: WordCount – uncertainty specification.

| Parameter | Values |
|---|---|
| spoutwait | {1, 2, 3, 4, 5, 6, 7, 8, 9, 10, 100, 1e3, 1e4} |
| splitters | {1, 2, 3, 4, 5, 6} |
| counters | {1, 2, 3, 4, 5, 6, 7, 8, 9, 10, 11, 12, 13, 14, 15, 16, 17, 18} |

Table 11: PCE results for WordCount.

| Performance | PCE | | | |
| Indices | Degree | Mean | SD | CoV |
| --- | --- | --- | --- | --- |
| Throughput | 3 | 2837.65 | 6323.81 | 222.85% |
| Latency | 2 | 3102.61 | 1698.86 | 54.76% |

While the results from the PCE method are very close to the real values, the accuracy is not as good as in the first two case studies. Table 11 shows the results from the PCE analysis. Similar to the first two case studies, the system shows different levels of robustness for different performance indicators. While the robustness of the system in terms of latency is relatively high (although it still might need improvement), the system suffers greatly with respect to throughput robustness.

Table 12: Accuracy of PCE with different number of samples for WordCount.

| Performance | PCE Samples | | | | MC Samples |
| Indices | 1000 | 700 | 300 | 100 | Required |
| --- | --- | --- | --- | --- | --- |
| Throughput | 3.2 E-02 | 1.1 E-01 | 1.8 E-01 | 3.0 E-01 | $> 1,000$ |
| Latency | 2.9 E-02 | 1.2 E-05 | 2.9 E-04 | 7.1 E-04 | $> 1,000$ |

As in the first two case studies, we measured the accuracy of the PCE method with different numbers of samples. The errors produced are shown in Table 12. Compared to the first two case studies, the accuracy is not as high, although the PCE method produces very low errors for latency (less than 1%). The distribution of throughput, on the other hand, proves difficult to estimate with low number of samples. Nevertheless, the PCE method was more than 97% accurate with 1,000 samples. The MC method, on the other hand, did not converge with 1,000 samples, as shown in the last column of Table 12.

The WordCount case study is quite different in the way the uncertain parameters are sampled. Here, the input parameters are assumed to have discrete values, and a factorial design of all input parameters is considered. In PCE, input parameters have to be modelled as one of the continuous probability distributions presented in Table 1. This may have an impact on how accurately these discrete values can be represented as distributions, and as a result, the accuracy of PCE estimations. Nevertheless, the PCE method performed reasonable well with 700 samples, with accuracy around 90%.

Table 13: Leave-one-out cross-validation errors of PCE with different levels of relative added noise (RAN) for WordCount.

| Performance | RAN level | | | |
| Indices | 20 | 10 | 5 | 1 |
| --- | --- | --- | --- | --- |
| Throughput | 4.60 E-02 | 4.60 E-02 | 4.60 E-02 | 4.70 E-02 |
| Latency | 2.30 E-02 | 2.30 E-02 | 2.30 E-02 | 2.30 E-02 |

Finally, the leave-one-out cross-validation errors of PCE with different RAN levels are shown in Table 13. The PCE approach performed extremely well when noise was introduced in the WordCount system, with errors less than 5% in the worst case. This indicates that the PCE approach is robust even when input parameters are discrete values.

## 8. Discussion

The experimental evaluation performed on the three case studies demonstrated the power of polynomial chaos expansion in estimating the distribution of performance indices given a set of uncertain parameters as input. Early performance estimates are important as they determine whether the software system meets the requirements, and allow for refactoring actions to be taken in time, thus to avoid expensive fixes.

Our approach nicely fits with the novel DevOps trend [1], where the operational data is used for guiding changes in software. New versions of a software system may be developed at a daily basis and performance engineers have to rely on incomplete and uncertain data to decide about the required changes. Our approach supports such decisions to be made before having the full information about the parameters under which the software system will perform.

The assumption of a distribution for the uncertain parameters represent a limitation of the approach, however a similar methodology was used in our previous work for performance and reliability evaluation [9, 13]. Some strategies can be used to mitigate this open issue, in fact the uncertainty of parameters is often influenced by the origin of components. Information from hardware manufactures, third party software vendors or system experts is useful in characterising the uncertainty of specific parameters. For instance, in some situations, the distribution of the source variables can be obtained and consequently, the desired parameter's distribution can be approximated from its own source.

Each case study used in the experimental evaluation was from a different stage of the software development lifetime, from software architecture design stage, where performance indices are obtained through model-based analysis, to an emulated environment, and finally, a real running system where performance is measured online. Results showed that, in general, the PCE method achieves high accuracy levels for all systems, especially when a high number of samples is used (1,000 samples), where the method is at least 97% accurate. Even with a low number of samples (100), the method performed very well in terms of accuracy in the majority of the cases. In the majority of the cases, the truncation degree was less than 7. The highest truncation degree resulted in the first case study (E-commerce system) where the response time for making a purchase required a polynomial of degree 16. Nevertheless, the PCE approach took only a few seconds to compute the coefficients.

The case that suffered the most by the decrease in the number of samples used for PCE was throughput measured in WordCount. One reason for this is the way the input parameters are modelled. We found that the PCE method works best when input parameters can be modelled as probability distributions. In WordCount, however, input parameters are discrete values, which impact the sampling process, and as a result the accuracy of PCE.

The PCE approach was compared to Monte Carlo simulations, where the sampling of input parameters is controlled using a dynamic stopping criterion. As the simulations are very costly, we set the upper bound on the number of samples to be used to 1,000. Results showed that the PCE approach outperformed the MC method, requiring significantly less samples to model performance robustness, saving up to 225 hours.

We also tested how the PCE method reacts to noise, by introducing different levels of Gaussian white noise to measurements. The signal-to-noise ratio was set at different levels, with higher RAN level indicating less discrepancy and noise. This ensures that the PCE method is reliable even when measurements are not very accurate. In general, the PCE was very robust and produced low errors even for low RAN. In WordCount on E-commerce, the accuracy of the method was above 97% for all noise levels. The results were slightly worse for JPetStore, where accuracy drops to around 80% for the lower RAN level, however the method is quite robust with accuracy above 92% in all other cases. Similar results are observed for the E-commerce system.

Apart form the estimation of the probability distribution, the PCE method provides a robustness measure for the performance indices, i.e., the coefficient of variance (CoV). The standard deviation and mean of a variable are expressed in the same units, so taking the ratio of these two allows the units to cancel, and results in the CoV, which is expressed as a percentage. While the mean estimates whether the software system meets performance requirements, CoV shows the extent of variability in relation to the mean of the sample. This ratio can then be compared to other such ratios in a meaningful way: the solution with the smaller CoV is less dispersed, and as a result more robust than the solution with the larger CoV, making us more confident regarding the performance of the system.

We observed that for the same system, while a certain performance index may be robust toward the changes in uncertain parameters values, other performance indices may suffer. This was, for instance, illustrated in Table 7 for the JPetStore system, where mean memory was very robust with low CoV (3.89%), while the other indices had high CoV values (up to around 53%). This indicates that the robustness of different performance indices may be orthogonal, hence in the case of refactoring actions required to fix the robustness of the system with respect to a certain performance index, its impact may be negative on the robustness of other indices. Multiobjective optimisation techniques [67, 68] could be used to address this issue, which is a priority for our future work.

## 9. Conclusion

This paper introduces a computationally-efficient approach for uncertainty propagation and robustness analysis of performance indices. The approach is based on the approximate representation of the performance indices using polynomial chaos expansions, and provides a qualitative and quantitative estimation of the effect of parameter uncertainties on performance results. The computational cost of PCE is significantly lower than the classical Monte Carlo method, saving up to 225 hours (improvement of 90%).

The application of the approach to three case studies showed that even with a relatively low-order approximation, the polynomial chaos expansions can accurately estimate the shape and tails of the output and states distribution for performance indices, providing a generally applicable approach for uncertainty propagation in robust performance estimation.

In conclusion, the PCE analysis method can be used to efficiently propagate the effect of uncertain parameters in software systems and calculate the robustness of performance indices. Additionally, since the PCE approach provides a means for accurate estimation of the shape of the distribution, it can form the basis of a procedure for applying refactoring techniques to shape the distribution, which would enhance the flexibility in addressing uncertainty compared to worst-case or minimum variance control.

**Acknowledgement**

This research was supported under Australian Research Council's Discovery Projects funding scheme, project number DE 140100017.

**References**

[1] A. Brunnert, A. van Hoorn, F. Willnecker, A. Danciu, W. Hasselbring, C. Heger, N. Herbst, P. Jamshidi, R. Jung, J. von Kistowski, A. Koziolek, J. Kroß, S. Spinner, C. Vögele, J. Walter, A. Wert, Performance-oriented DevOps: A research agenda, Tech. Rep. SPEC-RG-2015-01, SPEC Research Group — DevOps Performance Working Group, Standard Performance Evaluation Corporation (SPEC) (2015).
[2] M. A. Marsan, G. Balbo, G. Conte, S. Donatelli, G. Franceschinis, Modelling with generalized stochastic Petri nets, John Wiley & Sons, Inc., 1994.
[3] L. Kleinrock, Queueing systems, volume i: theory.
[4] G. Franks, D. C. Petriu, C. M. Woodside, J. Xu, P. Tregunno, Layered bottlenecks and their mitigation, in: Proceedings of the International Conference on the Quantitative Evaluation of Systems, 2006, pp. 103–114.
[5] M. Woodside, G. Franks, D. C. Petriu, The future of software performance engineering, in: Proceedings of the International Workshop on Future of Software Engineering, IEEE, 2007, pp. 171–187.
[6] N. Esfahani, S. Malek, K. Razavi, Guidearch: guiding the exploration of architectural solution space under uncertainty, in: Proceedings of the International Conference on Software Engineering, 2013, pp. 43–52.
[7] P. Jamshidi, C. Pahl, N. C. Mendonça, Managing uncertainty in autonomic cloud elasticity controllers, IEEE Cloud Computing 3 (3) (2016) 50–60.
[8] D. Garlan, Software engineering in an uncertain world, in: Proceedings of International Workshop on Future of Software Engineering Research, 2010, pp. 125–128.
[9] C. Trubiani, I. Meedeniya, V. Cortellessa, A. Aleti, L. Grunske, Model-based performance analysis of software architectures under uncertainty, in: Proceedings of the International Conference on Quality of Software Architectures, 2013, pp. 69–78.
[10] K. Sepahvand, S. Marburg, H.-J. Hardtke, Uncertainty quantification in stochastic systems using polynomial chaos expansion, International Journal of Applied Mechanics 2 (02) (2010) 305–353.
[11] D. C. Petriu, H. Shen, Applying the UML performance profile: Graph grammar-based derivation of LQN models from UML specifications, in: Proceedings of the International Conference on Modelling Techniques and Tools for Computer Performance Evaluation, Springer, 2002, pp. 159–177.
[12] I. Meedeniya, A. Aleti, L. Grunske, Architecture-driven reliability optimization with uncertain model parameters, Journal of Systems and Software 85 (10) (2012) 2340–2355.
[13] I. Meedeniya, I. Moser, A. Aleti, L. Grunske, Architecture-based reliability evaluation under uncertainty, in: Proceedings of the International Conference on Quality of Software Architectures, 2011, pp. 85–94.
[14] M. Awad, D. A. Menascé, On the predictive properties of performance models derived through input-output relationships, in: Proceedings of the European Workshop on Computer Performance Engineering, 2014, pp. 89–103.
[15] M. Awad, D. A. Menascé, Dynamic derivation of analytical performance models in autonomic computing environments, in: Proceedings of the Computer Measurement Group Conference, 2014.
[16] R. H. Myers, D. C. Montgomery, C. M. Anderson-Cook, Response surface methodology: process and product optimization using designed experiments, John Wiley & Sons, 2016.

[17] R. L. Mason, R. F. Gunst, J. L. Hess, Designs and analyses for fitting response surfaces, Statistical Design and Analysis of Experiments: With Applications to Engineering and Science, Second Edition (2003) 568–613.
[18] B. D. Youn, K. K. Choi, A new response surface methodology for reliability-based design optimization, Computers & structures 82 (2) (2004) 241–256.
[19] T. W. Simpson, J. J. Korte, T. M. Mauery, F. Mistree, Comparison of response surface and kriging models for multidisciplinary design optimization, NASA Technical Report, 1998.
[20] M. Farina, A neural network based generalized response surface multiobjective evolutionary algorithm, in: Proceedings of the International Congress on Evolutionary Computation, Vol. 1, IEEE, 2002, pp. 956–961.
[21] T. Goel, R. Vaidyanathan, R. T. Haftka, W. Shyy, N. V. Queipo, K. Tucker, Response surface approximation of pareto optimal front in multi-objective optimization, Computer methods in applied mechanics and engineering 196 (4) (2007) 879–893.
[22] Z. He, J. Wang, J. Oh, S. H. Park, Robust optimization for multiple responses using response surface methodology, Applied stochastic models in business and industry 26 (2) (2010) 157–171.
[23] J. C. Helton, J. D. Johnson, C. J. Sallaberry, C. B. Storlie, Survey of sampling-based methods for uncertainty and sensitivity analysis, Reliability Engineering & System Safety 91 (10) (2006) 1175–1209.
[24] Y. Jin, A comprehensive survey of fitness approximation in evolutionary computation, Soft Computing-A Fusion of Foundations, Methodologies and Applications 9 (1) (2005) 3–12.
[25] M. Courtois, C. M. Woodside, Using regression splines for software performance analysis, in: Proceedings of International Workshop on Software and Performance, 2000, pp. 105–114.
[26] S. Kraft, S. Pacheco-Sanchez, G. Casale, S. Dawson, Estimating service resource consumption from response time measurements, in: Proceedings of the International Conference on Performance Evaluation Methodologies and Tools, 2009, p. 48.
[27] W. Wang, G. Casale, Bayesian service demand estimation using gibbs sampling, in: International Symposium on Modelling, Analysis and Simulation of Computer and Telecommunication Systems, IEEE, 2013, pp. 567–576.
[28] A. Kattepur, M. Nambiar, Performance modeling of multi-tiered web applications with varying service demands, International Journal of Networking and Computing 6 (1) (2016) 64–86.
[29] N. Siegmund, S. S. Kolesnikov, C. Kästner, S. Apel, D. Batory, M. Rosenmüller, G. Saake, Predicting performance via automated feature-interaction detection, in: Proceedings of the International Conference on Software Engineering, IEEE, 2012, pp. 167–177.
[30] P. Jamshidi, M. Velez, C. Kästner, N. Siegmund, P. Kawthekar, Transfer learning for improving model predictions in highly configurable software, in: Proceedings of the International Symposium on Software Engineering for Adaptive and Self-Managing Systems, IEEE, 2017.
[31] J. Guo, K. Czarnecki, S. Apel, N. Siegmund, A. Wasowski, Variability-aware performance prediction: A statistical learning approach, in: Proceedings of the International Conference on Automated Software Engineering, IEEE, 2013, pp. 301–311.
[32] A. Sarkar, J. Guo, N. Siegmund, S. Apel, K. Czarnecki, Cost-efficient sampling for performance prediction of configurable systems, in: Proceedings of the International Conference on Automated Software Engineering, IEEE, 2015.
[33] N. Siegmund, A. Grebhahn, S. Apel, C. Kästner, Performance-influence models for highly configurable systems, in: Proceedings of the Joint Meeting on Foundations of Software Engineering, ACM, 2015, pp. 284–294.
[34] T. Zheng, C. M. Woodside, M. Litoiu, Performance model estimation and tracking using optimal filters, IEEE Transactions on Software Engineering 34 (3) (2008) 391–406.
[35] H. Ghanbari, C. Barna, M. Litoiu, M. Woodside, T. Zheng, J. Wong, G. Iszlai, Tracking adaptive performance models using dynamic clustering of user classes 36 (5) (2011) 179–188.
[36] B. C. Lee, D. M. Brooks, B. R. de Supinski, M. Schulz, K. Singh, S. A. McKee, Methods of inference and learning for performance modeling of parallel applications, in: Proceedings of Symposium on Principles and Practice of Parallel Programming, 2007, pp. 249–258.
[37] A. B. Sharma, R. Bhagwan, M. Choudhury, L. Golubchik, R. Govindan, G. M. Voelker, Automatic request categorization in internet services, ACM SIGMETRICS Performance Evaluation Review 36 (2) (2008) 16–25.
[38] M. Faber, J. Happe, Systematic adoption of genetic programming for deriving software performance curves, in: Proceedings of International Conference on Performance Engineering, ACM, 2012, pp. 33–44.
[39] N. Wiener, The homogeneous chaos, American Journal of Mathematics 60 (4) (1938) 897–936.

[40] D. Xiu, G. E. Karniadakis, The wiener–askey polynomial chaos for stochastic differential equations, SIAM journal on scientific computing 24 (2) (2002) 619–644.
[41] D. A. Menascé, H. Gomaa, S. Malek, J. P. Sousa, SASSY: A framework for self-architecting service-oriented systems, IEEE Software 28 (6) (2011) 78–85.
[42] M. Kowal, M. Tschaikowski, M. Tribastone, I. Schaefer, Scaling size and parameter spaces in variability-aware software performance models, in: Proceedings of the International Conference on Automated Software Engineering, IEEE, 2015, pp. 407–417.
[43] E. Incerto, M. Tribastone, C. Trubiani, Symbolic performance adaptation, in: Proceedings of the International Symposium on Software Engineering for Adaptive and Self-Managing Systems, ACM, 2016, pp. 140–150.
[44] D. Perez-Palacin, R. Mirandola, Dealing with uncertainties in the performance modelling of software systems, in: Proceedings of the International Conference on Quality of Software Architectures, 2014, pp. 33–42.
[45] P. Jamshidi, G. Casale, An uncertainty-aware approach to optimal configuration of stream processing systems., in: Proceedings of the IEEE International Symposium on the Modeling, Analysis, and Simulation of Computer and Telecommunication Systems, 2016.
URL http://dx.doi.org/10.5281/zenodo.56238
[46] P. Jamshidi, A. Sharifloo, C. Pahl, H. Arabnejad, A. Metzger, G. Estrada, Fuzzy self-learning controllers for elasticity management in dynamic cloud architectures, in: Proceedings of the International Conference on Quality of Software Architectures, IEEE, 2016, pp. 70–79.
[47] C. Pahl, P. Jamshidi, O. Zimmermann, Architectural principles for cloud software, ACM Transactions on Internet Technology.
[48] I. Meedeniya, A. Aleti, I. Avazpour, A. Amin, Robust archeopterix: Architecture optimization of embedded systems under uncertainty, in: Proceedings of the Second International Workshop on Software Engineering for Embedded Systems, IEEE Press, 2012, pp. 23–29.
[49] I. Meedeniya, I. Moser, A. Aleti, L. Grunske, Evaluating probabilistic models with uncertain model parameters, Software & Systems Modeling 13 (4) (2014) 1395–1415.
[50] M. Marseguerra, E. Zio, L. Podofillini, Multiobjective spare part allocation by means of genetic algorithms and monte carlo simulation, Reliability engineering & system safety 87 (3) (2005) 325–335.
[51] A. P. Dempster, N. M. Laird, D. B. Rubin, Maximum likelihood from incomplete data via the em algorithm, Journal of the royal statistical society. Series B (methodological) (1977) 1–38.
[52] D. C. Montgomery, G. C. Runger, Applied statistics and probability for engineers, John Wiley & Sons, 2010.
[53] L. Zadeh, Fuzzy sets, Information and Control 8 (3) (1965) 338 – 353.
[54] Y. Ben-Haim, Info-gap decision theory: decisions under severe uncertainty, Academic Press, 2006.
[55] R. E. Moore, R. B. Kearfott, M. J. Cloud, Introduction to interval analysis, Siam, 2009.
[56] A. Hyvärinen, J. Karhunen, E. Oja, Independent component analysis, Vol. 46, John Wiley & Sons, 2004.
[57] A. Keese, H. G. Matthies, Sparse quadrature as an alternative to monte carlo for stochastic finite element techniques, Proceedings in Applied Mathematics and Mechanics 3 (1) (2003) 493–494.
[58] T. Hastie, R. Tibshirani, J. Friedman, The elements of statistical learning: data mining, inference and prediction, 2nd Edition, Springer, 2008.
[59] G. Blatman, B. Sudret, An adaptive algorithm to build up sparse polynomial chaos expansions for stochastic finite element analysis, Probabilistic Engineering Mechanics 25 (2) (2010) 183–197.
[60] A. M. Molinaro, R. Simon, R. M. Pfeiffer, Prediction error estimation: a comparison of resampling methods, Bioinformatics 21 (15) (2005) 3301–3307.
[61] C. Trubiani, A. D. Marco, V. Cortellessa, N. Mani, D. C. Petriu, Exploring synergies between bottleneck analysis and performance antipatterns, in: Proceedings of the International Conference on Performance Engineering, 2014, pp. 75–86.
[62] G. Franks, P. Maly, M. Woodside, D. C. Petriu, A. Hubbard, M. Mroz, Layered Queueing Network Solver and Simulator, [online] http://www.sce.carleton.ca/rads/lqns/LQNSUserMan-jan13.pdf (2013).
[63] B. Hughes, On the error probability of signals in additive white gaussian noise, IEEE Transactions on Information Theory 37 (1) (1991) 151–155.
[64] A. van Hoorn, M. Rohr, W. Hasselbring, Generating probabilistic and intensity-varying workload for web-based software systems, in: Proceedings of International Performance Evaluation Workshop, 2008, pp. 124–143.
[65] B. White, J. Lepreau, L. Stoller, R. Ricci, S. Guruprasad, M. Newbold, M. Hibler, C. Barb,

A. Joglekar, An integrated experimental environment for distributed systems and networks, in: Proceedings of Symposium on Operating Systems Design and Implementation, 2002, pp. 255–270.
[66] P. Jamshidi, N. Siegmund, M. Velez, C. Kästner, A. Patel, Y. Agarwal, Transfer learning for performance modeling of configurable systems: An exploratory analysis, in: Proceedings of the International Conference on Automated Software Engineering, ACM, 2017.
[67] A. Aleti, I. Moser, S. Mostaghim, Adaptive range parameter control, in: Evolutionary Computation (CEC), 2012 IEEE Congress on, IEEE, 2012, pp. 1–8.
[68] A. Aleti, I. Moser, Predictive parameter control, in: Proceedings of the 13th annual conference on Genetic and evolutionary computation, ACM, 2011, pp. 561–568.